\documentclass[
 reprint,
 amsmath,amssymb,
pra
]{revtex4-1}

\usepackage[symbol]{footmisc}
\usepackage{graphicx}% Include figure files
\usepackage{dcolumn}% Align table columns on decimal point
\usepackage{bm}% bold math
\begin{document}

\title{MHz Gravitational Wave Constraints with Decameter Michelson Interferometers}
\def\andname{\hspace*{-0.3em}}
\author{Aaron S. Chou$^1$}
\author{Richard Gustafson$^2$ }
\author{Craig Hogan$^{1,3}$}
\author{Brittany Kamai$^{1,3,4}$}\thanks{Corresponding author : bkamai@caltech.edu}
\author{Ohkyung~Kwon$^{3,5}$}
\author{Robert Lanza$^{3,6}$}
\author{Shane L. Larson$^{7,8}$}
\author{Lee McCuller$^{3,6}$}
\author{Stephan~S.~Meyer$^{3}$}
\author{Jonathan Richardson$^{2}$}
\author{Chris Stoughton$^1$}
\author{Raymond Tomlin$^1$}
\author{Rainer Weiss$^6$}
\affiliation{\\ $^{1}$Fermi National Accelerator Laboratory;
$^{2}$University of Michigan;
$^{3}$University of Chicago; \\
$^{4}$Vanderbilt University;
$^{5}$Korea Advanced Institute of Science and Technology;\\
$^{6}$Massachusetts Institute of Technology;
$^{7}$Northwestern University; 
$^{8}$Adler Planetarium
}

\collaboration{The Holometer Collaboration}


\begin{abstract}
%1) What the instrument is
A new detector, the Fermilab Holometer, consists of separate yet identical 39-meter Michelson interferometers. Strain sensitivity achieved is better than $10^{-21} /{\sqrt{\rm{Hz}}}$ between 1 to 13 MHz from a 130-hr dataset. This measurement exceeds the sensitivity and frequency range made from previous high frequency gravitational wave experiments by many orders of magnitude. Constraints are placed on a stochastic background at 382 Hz resolution. The 3$\sigma$ upper limit on $\Omega_{\rm{GW}}$, the gravitational wave energy density normalized to the closure density, ranges from $5.6 \times 10^{12}$ at 1 MHz to $8.4 \times 10^{15}$ at 13 MHz. Another result from the same dataset is a search for nearby primordial black hole binaries (PBHB). There are no detectable monochromatic PBHBs in the mass range $0.83$ - $3.5 \times 10^{21}$g between the earth and the moon. Projections for a chirp search with the same dataset increases the mass range to $0.59 - 2.5 \times 10^{25}$g and distances out to Jupiter. This result presents a new method for placing limits on a poorly constrained mass range of primordial black holes. Additionally, solar system searches for PBHBs place limits on their contribution to the total dark matter fraction.

%\begin{description}
%For starting dance parties
%May be entered using the \verb+\pacs{#1}+ command to create ms. pacman
%\end{description}
\end{abstract}

\maketitle

%############################## INTRODUCTION ###############################
\section{Introduction}\label{sec:GWspectrum}
%word\footnote{sup}

Gravitational waves are predicted to exist at all frequencies and direct measurements from a broad range of 
experiments can probe a variety of sources. The strongest astrophysical sources radiate at frequencies less 
that a few kHz. Current experiments are either operating at or designed to search for sources at these frequencies \cite{LIGO2016_InstrumentLong,VIRGO2015, GEO600_2004,PPTA2010, EPTA2013,NANOGRAV2009,eLISA2014}. In this paper, we report a search for gravitational waves in a broad band from 1 MHz up to 13 MHz, using data from a new instrument, the Fermilab Holometer \cite{Holo2015}.  Our sensitivity exceeds earlier results on high frequency gravitational waves by orders of magnitude \cite{Bernard2001,Cruise2006}.

Potential sources at these high frequencies include an unresolved stochastic background from a super-position of many individual sources such as primordial black holes\cite{Carr1974a}, cosmic (super)-string loops \cite{Siemens2007} and other relics possibly produced in the early universe \cite{Cruise2012}. Additionally, those individual relics may still exist today and will be emitting gravitational radiation \cite{Carr1975}. Therefore, the data is analyzed here in two ways:  first as a constraint on a statistically isotropic stochastic background, and then as a constraint on individually resolved, non-chirping black hole binaries whose orbital frequencies lie in this frequency band. The mass range of black hole binaries probed in this search is $\sim 10^{21}$g with the potential to test up to $10^{26}$ g with the same dataset. This is one of the least constrained mass ranges for primordial black holes \cite{Carr2010}.

This paper begins with a description of the instrument, data acquisition system and data analysis pipeline. Next, the results for 
the stochastic gravitational wave background and the narrow-lined search for primordial black holes binaries are presented.

%=================== INSTRUMENT =======================
\section{The Holometer}\label{sec:experimental_design}

\subsection{The Instrument}\label{sec:instrument}

The Holometer is comprised of two identical power-recycled Michelson interferometers, separated by half a meter, with the same orientation in space. In each interferometer, a continuous wave 1064nm laser at the input is divided into two orthogonal paths by the beamsplitter, and sent down the 39-meter-long arms. Light returning from the end mirrors coherently interferes at the beamsplitter, where the constructively interfering light is resonantly enhanced by the power-recycling mirror at the interferometer's input. The returning destructive light exits the interferometer at the output of the interferometer where the arm length difference is measured as $\Delta L \equiv L_x - L_y$. The power-recycling technique increases the input laser power of 1W to aproximately 2 kW, which improves the shot-noise-limited displacement sensitivity of a single interferometer by an order of magnitude to $\sim2\times10^{-18}\,\mathrm{meters}/\sqrt{\mathrm{Hz}}$. 

By operating two identical interferometers separated by 0.635 m and cross-correlating their signal outputs, sub-shot-noise-limited performance is achieved. In order to minimize backgrounds from cross-talk, the interferometers are isolated optically, mechanically, and electronically. Each interferometer is equipped with separate lasers, electronics, and core optics (beamsplitter, power-recycling mirror, and two end mirrors) enclosed in independent ultra-high vacuum systems. Details of the setup 
appear in PhD theses \cite{Lanza2015,McCuller2015,Kamai2016,Richardson2016} and the Holometer instrument paper \cite{Holo2016_Instrument}. 

At MHz frequencies, photon shot noise is the dominant noise source, which is uncorrelated between the two interferometers. Therefore, cross-correlating the output of the interferometers increases the sensitivity as $1/\sqrt{N}$, for $N$ samples. This allows us to surpass the shot noise limited sensitivity of a single interferometer and hunt for signals far below the baseline shot noise level. Conversely, when a correlated signal is present in both detectors, the averaged cross-spectrum will converge to the level of the signal after a sufficient number of averages have been recorded. Extensive campaigns were conducted to verify that there are no unknown correlated noise sources above 1 MHz. Unknown correlated noise sources were characterized through extensive measurements before and after data taking campaigns. Additionally, laser phase and intensity noise was bounded using dedicated monitors during the runs.

\subsection{Dataset}\label{sec:obs}

The 130-hour dataset used in this analysis was collected from 15 July 2015 to 15 August 2015 at Fermi National Accelerator Laboratory. First, this dataset is analyzed to place constraints on the energy density of stochastic gravitational wave backgrounds. Then utilizing the same dataset, a different analysis is done to place constraints on primordial black hole pairs.

The design of the Holometer data acquisition system differs from other gravitational wave experiments such as Pulsar Timing Arrays (PTAs), Laser Interferometeric Gravitational-Wave Observatory (LIGO), and Laser Interferometric Space Antenna (LISA) \cite{LIGO2016_InstrumentLong,VIRGO2015, GEO600_2004,PPTA2010, EPTA2013,NANOGRAV2009,eLISA2014}. These other experiments continuously store time domain data to search for a range of sources from bursts to stochastic backgrounds. The Holometer was initially designed to look for a stationary noise source where storage of time domain data was unnecessary \cite{Holo2015}. Therefore, the data acquisition pipeline was written to retain only frequency domain data in the form of power- and cross- spectral densities for each channel. 

A total of 8 channels corresponding to the output of the interferometers along with environmental monitors are digitized at 100MHz sampling rate. To reduce the data rate per channel, two neighboring time-series measurements are averaged together to result in an effective 50 MHz sampling rate. Each $\sim$3-millisecond, a Fast-Fourier Transform (FFT) is calculated for each channel. Additionally, real-valued, power spectral density (PSD) is calculated for each channel and the complex-valued, cross-spectral density (CSD) is computed for each combination of channels (i.e 1x2, 1x3, 1x4, etc). Explicitly, the PSD is $|A_1|^2$ and the CSD is $A_1 A_2 e^{i(\theta_1 - \theta_2)}$ where $A$ is the amplitude of the Fourier transform, $\theta$ is the angle of the vector in the complex plane for channels $1$ and $2$. The total number of frequency bins between 0 and 13 MHz is 34,079, which has a frequency resolution of 382 Hz. After 1.4 seconds, 1,400 milli-second power- and cross-spectra are averaged together, GPS-time-stamped and recorded in the final hierarchical data formatted (\verb|hdf5|) file. 

Data vetoes were implemented to ensure that contaminated data (i.e. from large RFI spikes) were not included in the averaging. This procedure repeats until the end of data acquisition. The 1-second averaged spectra were calibrated against a 1kHz length dither to establish the conversion from V/Hz to m/Hz.

%====== Data Analysis Pipeline SGWB============
\section{Stochastic Gravitational Wave background}\label{sec:SGWB}

\subsection{Data Analysis pipeline}\label{sec:dataPipeSGWB}
\begin{figure}[h!]

\includegraphics[width=1.\linewidth]{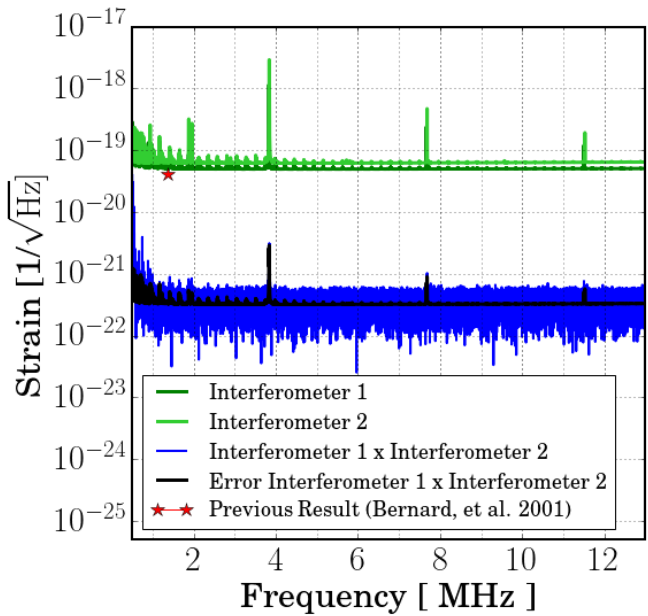}
\caption[Spectral density of 130-hour dataset]{Strain amplitude spectral density as a function of frequency for the 130-hour dataset from 0.5-13 MHz. The green traces are the power spectral densities for each of the interferometers (Interferometer 1 = dark green, Interferometermeter 2 = light green). The blue trace is the magnitude of the cross spectral density between Interferometer 1 and Interferometer 2. The black trace is the statistical uncertainty in the cross-spectral density. The spectral features above 500 kHz are well understood and described in Section~\ref{sec:dataPipeSGWB}.} \label{fig:SB_150HrPSD}
\end{figure}

To place constraints on the energy density of the stochastic gravitational wave background in the MHz frequency range, the entire 130-hour dataset was 
averaged to increase the strain sensitivity. The fully averaged strain spectral density is in Figure~\ref{fig:SB_150HrPSD}. Each interferometer's 
PSD is shown by the green traces, the CSD between both interferometers is shown by the blue trace, the error on the 
CSD is shown by the black trace. The 2 orders of magnitude gain in sensitivity between the power- and cross- spectral density are from averaging 
together complex numbers. The CSD error has the scaling as $\simeq \rm{\sqrt{PSD_1 PSD_2/N}} $ where N is the number of milli-second 
spectra used in the averaging. In practice, this was independently calculated from the sample variance of millisecond CSDs in 5-min averaged batches.
The final CSD error averages together the 5-min CSD errors, which was verified to directly follow the prediction 
of where the CSD should be given two Gaussian noise source (such as photon shot noise from each interferometer as measured by the PSDs) 
and the integration time (130 hours). Any excess measured by the blue trace above the black trace (i.e. what is shown below 1 MHz) 
is correlated noise between the two interferometers whereas greater than 1 MHz the noise that is consistent with the statistical error for uncorrelated noise.

All of the features above 500kHz in the individual traces of Figure~\ref{fig:SB_150HrPSD} are well understood. From 500kHz to 1 MHz, the dominant source of noise is laser phase and amplitude noise. From 1 MHz to 13 MHz, the dominant source of noise is photon shot noise. The large spikes at 3.75 MHz and its harmonics are due to laser noise leaking into each interferometer due to a lack of filtering from the Fabry-Perot cavity. Another type of noise is the clusters of spikes that begin in the low frequency end (most noticeable at $\sim$ 1 MHz) and have repeated decaying harmonics that are barely noticeable above 13 MHz. These are due to the drumhead modes of the optics in each interferometer. Studies have verified that each cluster of spikes actually consists of three spikes from the beamsplitter and two end mirrors. This noise source is independent for each interferometer and does not show up in the cross-correlated measurement.

The Holometer strain spectral density result surpasses the only other measurement previously made in this frequency range as shown by the red star in Figure~\ref{fig:SB_150HrPSD} \cite{Bernard2001}. Their measurement was performed using super-conducting microwave cavities with a narrow-line strain sensitivity of $3.3 \times 10^{-20}$ Hz$^{1/2}$  centered at 1.38 MHz with a bandwidth of 100 Hz.

\subsection{Result 1 : MHz Constraints on SGWB}
The energy density of a stochastic gravitational wave background is characterized by 
how the energy is distributed in frequency \cite{Allen1999}. This is parametrized by $\Omega_{gw}$ which relates the 
gravitational wave energy density in a bandwidth of $\Delta f$ to the total energy density to close the universe defined as \cite{Allen1999}

\begin{equation}
\Omega_{gw}(f) = \frac{1}{\rho_c} \frac{d\rho_{gw}}{d\,ln(f)}
\end{equation}

\noindent where $f$ is the frequency, $\rho_c$ is the closure (critical) energy density of the universe and $d\rho_c$ is the gravitational radiation energy 
density contained in the range from $f$ to $df$. 

% Useful to characterized the spectral properties of gravitational wave by specifying how the energy is distributed in frequency (Allen & Romano)
% isotropic, unpolarized, stationary & gaussian (Allen & Romano)

To compute the energy density from the strain measurements, $\hat{\Omega}_{gw}(f)$, the following relation is used 

\begin{equation}
\hat{\Omega}_{gw}(f) \equiv  \frac{\Re[h_{1,2}(f)]}{S(f)}
\end{equation}

\noindent where $\Re$ are the real components, $h_{1,2}(f)$ is the strain cross spectral density in [1/Hz] units and $S(f)$ is the conversion 
factor that is sky and polarization averaged, which is defined as 
\begin{equation}
S(f) = \frac{3H^2_0}{10\pi^2} \frac{1}{f^3} \label{eqn:strain}
\end{equation}    

\noindent where $H_0$ is the Hubble parameter (the expansion rate of the universe $H_0$ = 69 [km/s/Mpc]) \cite{Bennett2013}.
The noise on the measurement is calculated as 
\begin{equation}
\sigma^2_{\hat{\Omega}}(f) \approx \frac{\sigma^2_{1,2}(f)}{S^2(f)}
\end{equation}

\noindent that has a scaling relationship $ \approx \frac{\rm{PSD_{1}}\rm{PSD_{2}}}{\rm{N}}$ though it is calculated directly from the averaged CSDs as discussed in Section.~\ref{sec:dataPipeSGWB}.

\begin{figure}[htb!]
\includegraphics[width=1.0\linewidth]{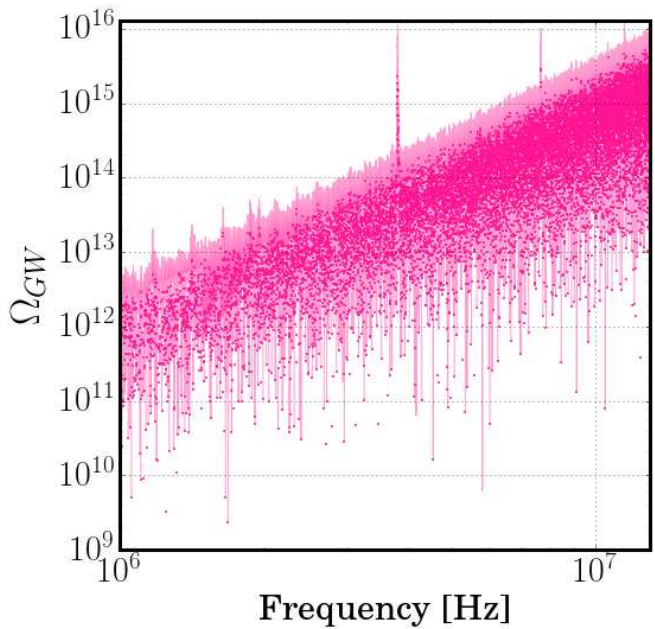}
\caption[Holometer Constraints on the Stochastic Gravitational Background] {\label{fig:OmegaGW_holo} Experimental constraints on the energy density of the gravitational wave backgrounds,  $\Omega_{\rm{GW}}$, as function of frequency. Each dot represents the measurement for a single 382 Hz frequency bin 
and the shaded region represents the 3$\sigma$ upper limit on one of the frequency bins. This is the first direct measurement of the 
stochastic gravitational backgrounds at these frequencies. This measurement is higher than both the closure density of the universe 
($\Omega_{\rm{GW}}$=1) and indirect limits set by the cosmic microwave background and the big bang nucleosynthesis which are 
at $\Omega_{\rm{GW}} \approx 10^{-5}$. In order to be competitive with these other probes, a major overhaul to the Holometer's sensitivity of the individual interferometers would need to be achieved. } 
\end{figure}

The measurement of the gravitational wave energy density, $\Omega_{gw}$, is plotted in Figure~\ref{fig:OmegaGW_holo}. Each individual point represents the value of the energy density for each 382 Hz frequency bin where the shaded region illustrates the corresponding 3$\sigma$ values. The 3$\sigma$ value of the energy density at 1 MHz is $5.6 \times 10^{12}$ and goes up to $8.4 \times 10^{15}$ at 13 MHz. This result has an additional 15\% systematic error from calibration uncertainties. It is should be noted that each frequency bin has a 50\% overlap fraction with the neighboring bin from the Hanning-Window function used in the FFT computation, which is important for the direct interpretation of the $\Omega_{gw}$ value for each frequency bin. Additionally, this result does not account for the degradation in sensitivity beyond the long wavelength approximation \cite{Schutz2011}, which is applicable to frequencies above 1.92 MHz, and would need to be properly accounted for as done for space-based gravitational wave missions. \cite{Larson2000}

This is the only direct measurement at MHz frequencies of the energy density of a stochastic gravitational wave background. This result is much higher than the closure density that would have a value of 1 in these units. Additionally, it is higher than indirect measurements at these frequencies from integrated limits placed by the Cosmic Microwave Background \cite{Sendra2012} and Big Bang Nucleosynthesis\cite{Cyburt2005} that have limits of $\sim10^{-5}$ in these units. At the current sensitivity, continuous integration would not be the right solution because it would take $10^{24}$ times longer than the current age of the universe to reach the closure density. In order to improve this measurement, a major overhaul to the instrument design must be done to increase its sensitivity. Since the Holometer is shot noise limited at these 
frequencies, the efforts must be invested in increasing each interferometer's power.

% Note to self : I never say why I chose the 1-13 MHz frequency range

%========== Data Analysis Pipeline : Narrow-lined Sources ===============

\section{Primordial Black Hole Binaries}

\subsection{Data analysis pipeline}\label{sec:dataPipenarrow}
Using the same 130-hour dataset, a different analysis was performed to search for stationary, monochromatic gravitational wave sources. 
This search evaluated individual frequency bins as a function of position on the sky over the duration of data acquisition. If a real monochromatic 
source exists, the strain signal will be the highest at some RA then as the earth rotates away the signal will decay to become consistent with 
the noise. If there is no amplitude decay pattern, then the excess strain power identified in some RA bin is excluded as a gravitational wave candidate. Time domain template matching is not possible since the Holometer data acquisition system only retains spectral density measurements.

\begin{figure}[h!]
\centering
\includegraphics[width=1.0\linewidth]{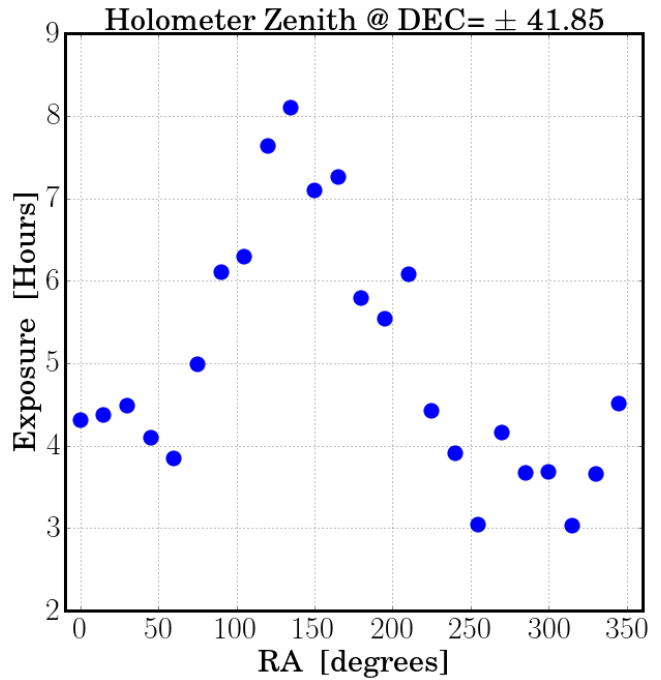}
\caption[Sky Exposure of 130-hour Dataset]{\label{fig:Holo_Exposure} The exposure as a function of RA for the Holometer zenith and anti-zenith for a declination of 
+41.85. The 130-hour dataset was split up into 24 RA bins and each dot represents the exposure time in each RA bin. The modulation in the exposure time is only representative of the amount of available operators. The minimum of $\sim$ 3 hours at RA $\sim$ 260 corresponded to midnight while the maximum at 8.75 hours corresponded to $\sim$ 4pm CST. } 
\end{figure}

First, the 130-hour dataset was sorted into 24 RA bins as shown in Figure~\ref{fig:Holo_Exposure}. Each blue point represents the amount of 
exposure for each RA bin of the Holometer's zenith and anti-zenith. The variation in exposure time was dependent on operator 
availability. Next, the average of power- and cross- spectral densities is calculated per RA bin, which results in a total of 24 RA binned spectra. 

The frequency range used in this narrow-lined search was limited to 1 - 1.92 MHz. The low frequency cut-off avoids correlated laser noise and high frequency cut-off is the valid limit for the long wavelength approximation, which other large scale interferometers such as aLIGO use \cite{Weiss1972, LIGO2016_instrument,LIGO2016_InstrumentLong}. Additionally, this search is for monochromatic gravitational wave sources, which means the 
gravitational wave frequency must not change by more than 382 Hz (frequency bin width) during the course of a month of observation taking. 

In the design and implementation of the Holometer, the interferometers have been verified to be perfectly in phase across this frequency range \cite{Holo2015, Holo2016_Instrument}. Therefore, the real component of the cross-spectral density for each frequency bin was used to search for a gravitational wave signal. The imaginary component was used as an independent measure of the noise distribution. 

A signal-to-noise was computed using the real component of the CSD to the error for each of the 2,396 frequency bins between 1-1.92 MHz. This signal-to-noise ratio was computed for each of the 24 RA binned spectra. The 57,504 signal-to-noise ratios were verified to be consistent with 
a Gaussian distribution. For comparison, the same signal-to-noise ratio was calculated using the imaginary component rather than the real component. This was also consistent with a Gaussian distribution, which verifies that the noise model is well understood. 

A gravitational wave candidate could still exist that would be consistent with a Gaussian distribution. Any frequency bins with a signal-to-noise value higher than 4 were flagged as potential candidates and a total of two potential candidates were followed up individually. For each of the potential candidates, the real component of the CSD was plotted as a function of RA. These potential candidates had some high 
CSD value at only a single RA bin but were consistent with zero in the neighboring ones, which is consistent with noise. The 
behavior was the same when compared to the imaginary component of the CSD as a function of RA. Using this method, the 
existence of a stationary astrophysical, narrow-lined source is ruled-out.

\subsection{Result 2 : PBH Constraints}\label{sec:PBHresults}

In the data analysis pipeline described above, no narrow-lined sources were found. This result is used to place constraints on primordial 
black hole binaries. Chirp masses and distances are computed based on the frequency range and strain sensitivity. 
This search was defined to look for monochromatic sources, meaning that if there were an 
inspiralling binary pair there would be no detectable change in frequency during the duration of data acquisition. 

The chirp mass, $M_c$, of the 
binary system can be calculated given this constraint on the change in gravitational wave frequency ($\Delta f_{\rm{gw}} = 382$ Hz), the observing time ($\Delta t = 1$ month) and the individual frequency bin ($f_{\rm{gw}}$) \cite{Peters1963,Larson2001}:

\begin{equation}\label{eqn:mass}
M_c =  (\alpha \,f_{gw}^{-11/3} \frac{\Delta f_{gw}}{\Delta t})^{3/5}
\end{equation}

\noindent where $ \alpha$ is $\frac{5}{96} \, \frac{1}{\pi^{8/3}}\, (\frac{c}{G^{1/3}})^5 $, $c$ is the speed of light and $G$ is the Newtonian gravitational constant. 
Given the frequency range of 1 - 1.92 MHz,  the range of chirp masses are $8.3\times 10^{20} -  3.5 \times 10^{21}$ g.  The distance to these binary pairs is computed in the following way \cite{Peters1963,Larson2001}: 

\begin{equation}\label{eqn:distance}
D = \frac{\beta}{\rm{SNR}} M_{c}(\pi f_{gw} M_{c})^{2/3} \frac{1}{h^{det}_{f_{gw}}} \sqrt{N}
\end{equation}

\noindent where $\beta$ is the combination of constants $\frac{G^{5/3}}{c^4}  (\frac{\pi^{2}}{2})^{1/3}$ , $c$ is the speed of light, $G$ is the Newtonian gravitational constant, $\rm{SNR}$ is the signal-to-noise ratio, $N$ is the number of cross spectral densities used in the averaging and $h^{det}_{f_{gw}}$ is the instantaneous strain of the detector at the frequency $f_{gw}$. Measurements were repeatedly taken on all parts of the sky and the RA bin with the longest exposure was used to calculate the distance. 

\begin{figure}[h!]
\centering
\includegraphics[width=1.0\linewidth]{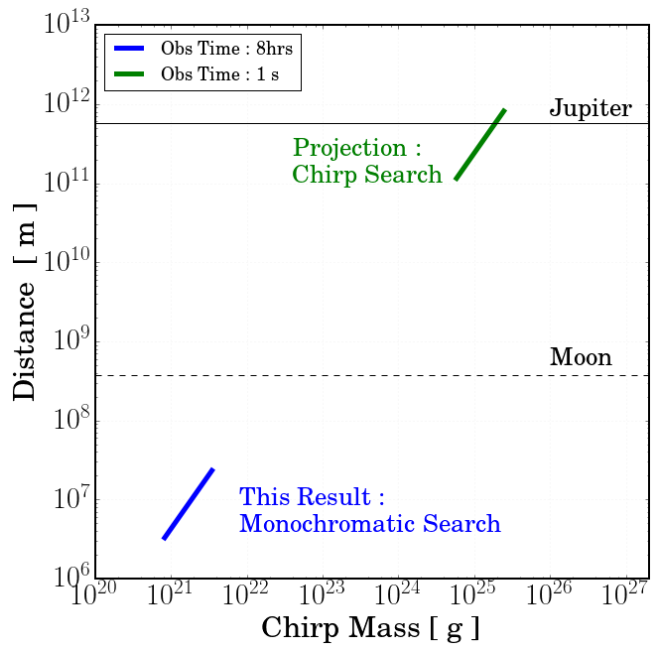}
\caption[Distance for Primordial Black Hole Masses with Alternative Analysis]{\label{fig:HorizonDistancePBH} Primordial black hole distance as a function of 
chirp mass. This is a comparison of the distance for this monochromatic frequency analysis (blue) and a projection for merging frequency stacked analysis (green). 
For reference, the dashed black line shows the distance to the moon, the solid black line shows the distance to Jupiter. }
\end{figure}

Figure~\ref{fig:HorizonDistancePBH} is the distance as a function of chirp mass where $\rho$ = 4, $h^{det}_{f_{gw}} = 5 \times 10^{-21} \, [1/\sqrt{\rm{Hz}}]$ 
and $N \approx 1.6 \times 10^{5}$ for $\sim$8 hrs of integration. The blue trace corresponds to the distance and the chirp masses calculated in this analysis. For reference, the distance to the moon (dashed trace) is shown. 

We report a null result of primordial black hole binaries with masses $8.3\times 10^{20} -  3.48 \times 10^{21}$~g between the earth and the moon. This is the first constraint of primordial black hole binaries in this distance range. This is the most conservative estimate for PBH pairs that can be tested with this dataset and an alternative analysis path will be presented below.

%=========== FUTURE WORK ===============

\section{Analysis Opportunities}\label{sec:futurework}

The analysis above for MHz gravitational wave sources is far from exhaustive. Searches for more massive primordial black holes binaries, 
individual cosmic strings, or the collision of vacuum bubbles from early universe phase transitions
or inflation is possible. New analysis with this dataset and/or data from identical operating conditions and storing the time series data (rather than on the averaged power- and cross- spectral densities) would make additional tests of these ideas possible. 

A search for more massive primordial black hole binary pairs is possible with this same dataset. The difference would be a frequency stacked search for chirping black hole binaries rather than the monochromatic black hole binaries searched for in this paper. The new analysis would use the 1-second averaged (1,400 millisecond) spectra, the shortest time averaged spectra stored to disk from this observing run. New chirp masses are calculated from equation \ref{eqn:mass} using the instantaneous sensitivity of $5 \times 10^{-21} \, /\sqrt{\rm{Hz}}$ and the chirp condition of $\Delta f / \Delta t > 382 \rm{\,Hz}/1 \rm{\, s}$ (rather than $\Delta f / \Delta t > 382 \rm{\,Hz}/1 \rm{\, \rm{month}}$ used in the monochromatic search). This would increase the chirp masses by 4 orders of magnitude and the increase the distance out to Jupiter as shown by the green trace in Figure~\ref{fig:HorizonDistancePBH}. Further details of all the above mentioned analysis can be found in Kamai Ph.D. dissertation \cite{Kamai2016}.

\section{Conclusions}

In this paper, we have demonstrated how decameter scale Michelson interferometers can be used for MHz gravitational wave 
searches. Employing the Fermilab Holometer, dual power-recycled 39 m Michelson Interferometers, strain sensitivities better than 
$10^{-21} \, [1/\sqrt{\rm{Hz}}]$ are achieved with a 130-hr dataset obtained between July to August, 2015. This sensitivity spans 
from 1 to 13 MHz with a frequency resolution of 382 Hz, which surpasses previous measurements both in 
strain sensitivity and frequency range as shown in Figure \ref{fig:SB_150HrPSD}. %Extensive campaigns were conducted to verify that the dominant source of nose 

The first gravitational wave measurement with this dataset is a constraint on the energy density of gravitational waves from 
a statistically, isotropic stochastic gravitational wave background. The 3$\sigma$ upper limit on the energy density, $\Omega_{\rm{gw}}$, is $5.6 \times 10^{12}$ at 1 MHz and goes up to $8.4 \times 10^{15}$ at 13 MHz as shown in Figure~\ref{fig:OmegaGW_holo}. This constraint places a direct upper bound 
on the contribution of gravitational waves to the total energy budget in each 382 Hz frequency bin. This limit is much higher than the closure 
density in a $\lambda$-CDM universe and indirect measurements (such as CMB and BBN), however this is the first direct measurement 
in this frequency range.

The second gravitational wave measurement with this dataset is a constraint on $0.85$ to $3.5 \times 10^{21}$g primordial black holes binaries (PBHB). 
The search was defined to look for stationary, monochromatic sources (i.e. no detectable change in frequency over the range of data acquisition). 
No frequency bins between 1 to 1.92 MHz with a signal-to-noise threshold of 4 exhibited a modulation pattern consistent with an antenna pattern.
Therefore, we exclude the existence of monochromatic PBHBs between the earth to the moon as shown by the blue trace in 
Figure~\ref{fig:HorizonDistancePBH}.

Projections are given for doing a chirp search with the same dataset, which can extend the PBHB mass range 
search up to $0.59 - 2.5 \times 10^{25}$g. This new mass range increases the distance range out to Jupiter as shown by the green trace in 
Figure~\ref{fig:HorizonDistancePBH}. Solar system searches can effectively constrain the dark matter contribution from light PBHs. These measurements are a new way to search for primordial black holes in one of the least constrained mass ranges ($10^{20}$~to~$10^{26}$g) \cite{Carr2010}. The Holometer or a Holometer-like experiment opens up a new opportunity to improve measurements in this mass range.\\
\\

This work was supported by the Department of Energy at Fermilab under Contract No. DE-AC02-07CH11359 and the Early Career Research Program (FNAL FWP 11-03), and by grants from the John Templeton Foundation, the National Science Foundation (Grants No. PHY-1205254 and No. DGE-1144082), NASA (Grant No. NNX09AR38G), the Fermi Research Alliance, the Kavli Institute for Cosmological Physics, University of Chicago/Fermilab Strategic Collaborative Initiatives, Science Support Consortium, and the Universities Research Association Visiting Scholars Program. B.K. was supported by National Science Foundation Graduate Research Fellowship Program (DGE-0909667), Universities Research Association Visiting Scholars Program and the Ford Foundation. O.K. was supported by the Basic Science Research Program (Grant No. NRF-2016R1D1A1B03934333) of the National Research Foundation of Korea (NRF) funded by the Ministry of Education. L.M. was supported by National Science Foundation Graduate Research Fellowship Program (DGE-0638477). The Holometer team gratefully acknowledges the extensive support and contributions of Bradford Boonstra, Benjamin Brubaker, Andrea Bryant, Marcin Burdzy, Herman Cease, Tim Cunneen, Steve Dixon, Bill Dymond, Valera Frolov, Jose Gallegos, Hank Glass, Emily Griffith, Hartmut Grote, Gaston Gutierrez, Evan Hall, Sten Hansen, Young-Kee Kim, Mark Kozlovsky, Dan Lambert, Scott McCormick, Erik Ramberg, Doug Rudd, Geoffrey Schmit, Alex Sippel, Jason Steffen, Sali Sylejmani, David Tanner, Jim Volk, Sam Waldman, William Wester, and James Williams for the design and construction of the apparatus.

\bibliographystyle{ieeetr}
\bibliography{Bibliography_MHzGW.bib}

\begin{thebibliography}{10}

\bibitem{LIGO2016_InstrumentLong}
D.V.~{Martynov} {\it et al.} (LIGO Scientific Collaboration), {arXiv:1604.00439}.

\bibitem{VIRGO2015}
F.~Acernese {\it et al.} (Virgo Collaboration), Class. Quantum Gravity, {\bf~32}, ~2, (2015).

\bibitem{GEO600_2004} 
H.~Grote {\it et al.} (GEO 600 Collaboration), ``{The status of GEO 600},' Class. Quantum Gravity, {\bf 22}, 10, (2004).

\bibitem{PPTA2010}
G.~Hobbs {\it et al.} (IPTA Collaboration),  Class. Quantum Gravity {\bf 27},  ~27, (2010).

\bibitem{EPTA2013}
M.~Kramer and D.~J.~Champion, {Class. Quantum Gravity}, {\bf~30},~22, (2013).

\bibitem{NANOGRAV2009}
F.~Jenet, {\it et al.} (NanoGrav Collaboration), arXiv:0909.1058

\bibitem{eLISA2014}
S.~Vitale, {Gen. Relativ. Gravit.}, {\bf 46}, (2014).

\bibitem{Holo2015}
A.~S. Chou,  {\it et al.} (Holometer Collaboration) {Phys. Rev. Lett.}, {\bf 117}, 111102 (2016).

\bibitem{Bernard2001}
P.~Bernard, G.~Gemme, R.~Parodi, and E.~Picasso, {Rev. Sci. Instrum.}, {\bf ~72}, ~5, (2001).

\bibitem{Cruise2006}
A.~M. Cruise and R.~M.~J. Ingley, {Class. Quantum Gravity}, {\bf~23}, 22, 2006.

\bibitem{Carr1974a}
B.~J.~Carr and S.~Hawking, {Mon. Not. R. Astron. Soc.}, {\bf 168}, (1974).

\bibitem{Siemens2007}
X.~Siemens, V.~Mandic, and J.~Creighton, {Phys. Rev. Lett.}, {\bf~98}, ~11, (2007).

\bibitem{Cruise2012}
A.~M. Cruise, {Class. Quantum Gravity}, {\bf ~29},~9, (2012).

\bibitem{Carr1975}
B.~J. Carr, {Astrophys. J.}, {\bf~201}, (1975).

\bibitem{Carr2010}
B.~J. Carr, K.~Kohri, Y.~Sendouda, and J.~Yokoyama, {Phys. Rev. D}, {\bf~81}, 10, (2010).

\bibitem{Allen1999}
B.~Allen and J.~D. Romano, {Phys. Rev. D}, {\bf~59}, ~10, (1999).

\bibitem{Bennett2013}
C.~L. Bennett, {\it et al.}, {Astrophys. J. Suppl. Ser.}, {\bf~208}, 2, (2013).

\bibitem{Schutz2011}
B.~F. Schutz, {Class. Quantum Gravity}, {\bf ~28}, ~12, (2011).

\bibitem{Larson2000}
S.~L. Larson, W.~Hiscock, and R.~Hellings, {Phys. Rev. D}, {\bf~62}, 6, (2000).

\bibitem{Sendra2012}
I.~Sendra and T.~L. Smith, {Phys. Rev. D }, {\bf~85}, ~12, (2012).

\bibitem{Cyburt2005}
R.~H. Cyburt, B.~D. Fields, K.~A. Olive, and E.~Skillman, {Astropart. Phys.}, {\bf~23}, ~3, (2005).

\bibitem{Weiss1972}
R.~Weiss, {Quarterly Report of the Research Laboratory for Electronics, MIT}, 1972.

\bibitem{LIGO2016_instrument}
B.P. Abbott {\it et al.} (LIGO Scientific Collaboration) and (Virgo Collaboration), {arxiv:1602.03838}

\bibitem{Peters1963}
P.~Peters and J.~Mathews,{Phys. Rev.},{\bf 131}, 1, (1963).

\bibitem{Larson2001}
S.~L. Larson, 2001, (unpublished)

\bibitem{Battaglia2005}
G.~Battaglia, A.~Helmi, H.~Morrison, P.~Harding, E.~W. Olszewski, M.~Mateo,
  K.~C. Freeman, J.~Norris, and S.~A. Shectman, {\em Mon. Not. R. Astron Soc.}, {\bf 364}, 2, (2005).
  
\bibitem{Lanza2015}
R.~K.~Lanza, Ph.D. thesis, University of Chicago, 2015.  
  
\bibitem{McCuller2015}
L.~McCuller, Ph.D. thesis, University of Chicago, 2015.
  
\bibitem{Kamai2016}
B.~Kamai, Ph.D. thesis, Vanderbilt University, 2016.

\bibitem{Richardson2016}
J.~Richardson, Ph.D. thesis, University of Chicago, 2016.

\end{thebibliography}

\end{document}